\documentclass[apj]{emulateapj}

\newcommand{\OIII}{\mbox{O\,\textsc{iii}}}

\newcommand{\OI}{\mbox{O\,\textsc{i}}}
\newcommand{\NII}{\mbox{N\,\textsc{ii}}}

\newcommand{\kms}{km s$^{-1}$}
\newcommand{\Ha}{H$\alpha$}

\slugcomment{accepted to ApJ}

\shorttitle{A Census of Gas Outflows in Type 2 AGNs}
\shortauthors{Bae \& Woo}

\begin{document}

\title{A Census of Gas Outflows in Type 2 Active galactic nuclei}

\author{Hyun-Jin Bae$^{1}$}
\author{Jong-Hak Woo$^{2,}$\altaffilmark{3,4}}

\affil{$^{1}$Department of Astronomy and Center for Galaxy Evolution Research, Yonsei University, Seoul 120-749, Republic of Korea; hjbae@galaxy.yonsei.ac.kr} 
\affil{$^{2}$Astronomy Program, Department of Physics and Astronomy, Seoul National University, Seoul 151-742, Republic of Korea; woo@astro.snu.ac.kr}
\altaffiltext{3}{TJ Park Science Fellow}
\altaffiltext{4}{Author to whom any correspondence should be addressed}

\begin{abstract}
We perform a census of ionized gas outflows 
using a sample of $\sim$23,000 type 2 active galactic nuclei (AGNs) out to z$\sim$0.1.
By measuring the velocity offset of narrow emission lines, i.e., [\OIII] $\lambda$5007 and 
\Ha, with respect to the systemic velocity measured from the stellar absorption lines, 
we find that 47\% of AGNs display an [\OIII] line-of-sight velocity offset $\geq$ 20 \kms. The fraction of the [\OIII] velocity offset in type 2 AGNs is
comparable to that in type 1 AGNs after considering the projection effect. 
AGNs with a large [\OIII] velocity offset preferentially have a high Eddington ratio, implying that the detected velocity offsets are related to black hole activity. 
The distribution of the host galaxy inclination is clearly different between the AGNs with
blueshifted [\OIII] and the AGNs with redshifted [\OIII], supporting the combined model of the biconical outflow
and dust obscuration. 
In addition, for $\sim$3\% of AGNs, [\OIII] and H$\alpha$ show comparable large velocity offsets,
indicating a more complex gas kinematics than decelerating outflows
in a stratified narrow-line region.
\end{abstract}

\keywords{galaxies: active --- galaxies: kinematics and dynamics}

\section{Introduction}
Supermassive black holes (SMBHs) and their host galaxies show relatively tight scaling relationships in the local Universe \citep[e.g.,][]{fm00,ge00,gu09,mm13,kh13, woo13}, implying the coevolution of black holes and their host galaxies. While theoretical studies suggested that mechanical and/or radiative feedback from active galactic nuclei (AGNs) regulate the growth of their host galaxies \citep[e.g.,][]{cr06,ci09,ps13,sh13,du13}, gas outflows detected in quasars may indicate the AGN feedback in action \citep[e.g.,][]{gr12}. Since AGN-driven outflows can significantly influence the surrounding interstellar medium (ISM) and star formation, observational constraints of the outflows are of importance in understanding SMBH-galaxy coevolution as well as the AGN feedback.

One of the tracers of gas outflows in AGNs is the line-of-sight velocity offset ($v_{\rm{off}}$) of narrow emission lines with respect to the systemic velocity of host galaxies \citep[e.g.,][]{bo05,ba09,ms11}. In particular, the velocity offset of the strong [\OIII] $\lambda$5007 narrow emission-line has long been of interest for probing AGN-driven outflows \citep[e.g.,][]{ck00,za02,bo05,ko08,cr10,zh11}. A number of high spatial resolution studies on nearby Seyfert galaxies reported decelerating radial gas motion in the narrow-line region (NLR), which were generally interpreted as AGN-driven 
biconical outflows \citep[e.g.,][]{ck00,cr00,ru01,ru05,da05,fi10,fi13}. 
Using Seyfert 1 galaxies and quasars, several groups have performed statistical investigations on the velocity offset of [\OIII] based on single-aperture spectra \citep[e.g.,][]{bo05,ko08,cr10,zh11}. For example, \citet{bo05} studied the [\OIII] velocity offset of $\sim$400 quasars selected from Sloan Digital Sky Survey (SDSS), suggesting that the radial velocity of [\OIII]-emitting gas is governed by both black hole mass and the Eddington ratio. \citet{ko08} showed that the [\OIII] lines of narrow-line Seyfert 1 galaxies are blueshifted with respect to the low-ionization lines, which is consistent with the 
decelerating outflows in the stratified NLR \citep[e.g.,][]{ck00, ru05}.
In this scenario inner high-ionization lines, e.g., [\OIII], show large radial velocity while outer low-ionization lines show no or smaller radial
velocity \citep{ko08}. For type 1 AGNs, however, it is not clear whether 
low-ionization lines are also shifted with respect to the systemic velocity, 
since the systemic velocity is not generally measured, for example, 
from stellar absorption lines.

For type 2 AGN, in contrast, the velocity offsets of both high- {\it and} low-ionization lines 
can be simultaneously measured with respect to the systemic velocity of their host galaxies
using the stellar absorption lines. 
Nevertheless, AGN-driven outflows based on the velocity offset 
measurements have not been extensively studied for type 2 AGNs while
there are several studies on the [\OIII] velocity offset with various limitations: 
some studies focused on a small sample \citep[e.g.,][]{co09,cr10},  or 
individual objects \citep[e.g.,][]{ck00,su05,ba12}, and other studies used low-ionization lines 
to infer the systemic velocity \citep[e.g.,][]{wa11}.
Interestingly, the results from these previous studies indicate the diverse origins of the 
[\OIII] velocity offsets. 
For a sample of 30 Seyfert 2 galaxies in DEEP2 survey, \citet{co09} reported similar velocity offsets between the [\OIII] 
and the H$\beta$ lines with respect to the stellar absorption lines, claiming that the velocity offsets are 
due to an inspiralling SMBH in merger remnants. Using a sample of 40 local Seyfert 2 galaxies, 
in contrast, \citet{cr10} reported that the velocity offset of [\OIII] with respect to stellar lines 
is due to a combination of biconical outflows and dust extinction. Meanwhile, \citet{wa11} 
based on a large sample of $\sim$3,000 SDSS type 2 AGNs argued that the velocity offset is not correlated with the Eddington ratio, which is in contrast to the results of the previous studies on
type 1 AGNs \citep[e.g.,][]{bo05,ko08}.

To investigate the statistical properties of gas outflows in type 2 AGNs, we perform a 
census of AGN-driven outflows by uniformly measuring the velocity offsets of narrow emission lines, i.e., [\OIII] 
and H$\alpha$, with respect to the stellar absorption lines, 
using a large sample of type 2 AGNs selected from SDSS Data Release 7 \citep[DR 7,][]{ab09}.
The paper is constructed as follows. In Section 2, we describe our sample selection and analysis methods. Section 3 presents the main results on the velocity offsets. In Section 4, we compare our results with those of previous studies, and discuss the origin of the detected velocity offsets. Section 5 summarizes our findings. In the paper we adopt $\Lambda$CDM cosmological parameters, i.e., $H_{o}$ = 70 \kms\ Mpc$^{-1}$ , $\Omega_{\Lambda}$ = 0.73, and $\Omega_{m}$ = 0.27.   

\begin{figure*}
\centering
\includegraphics[width=0.8\textwidth]{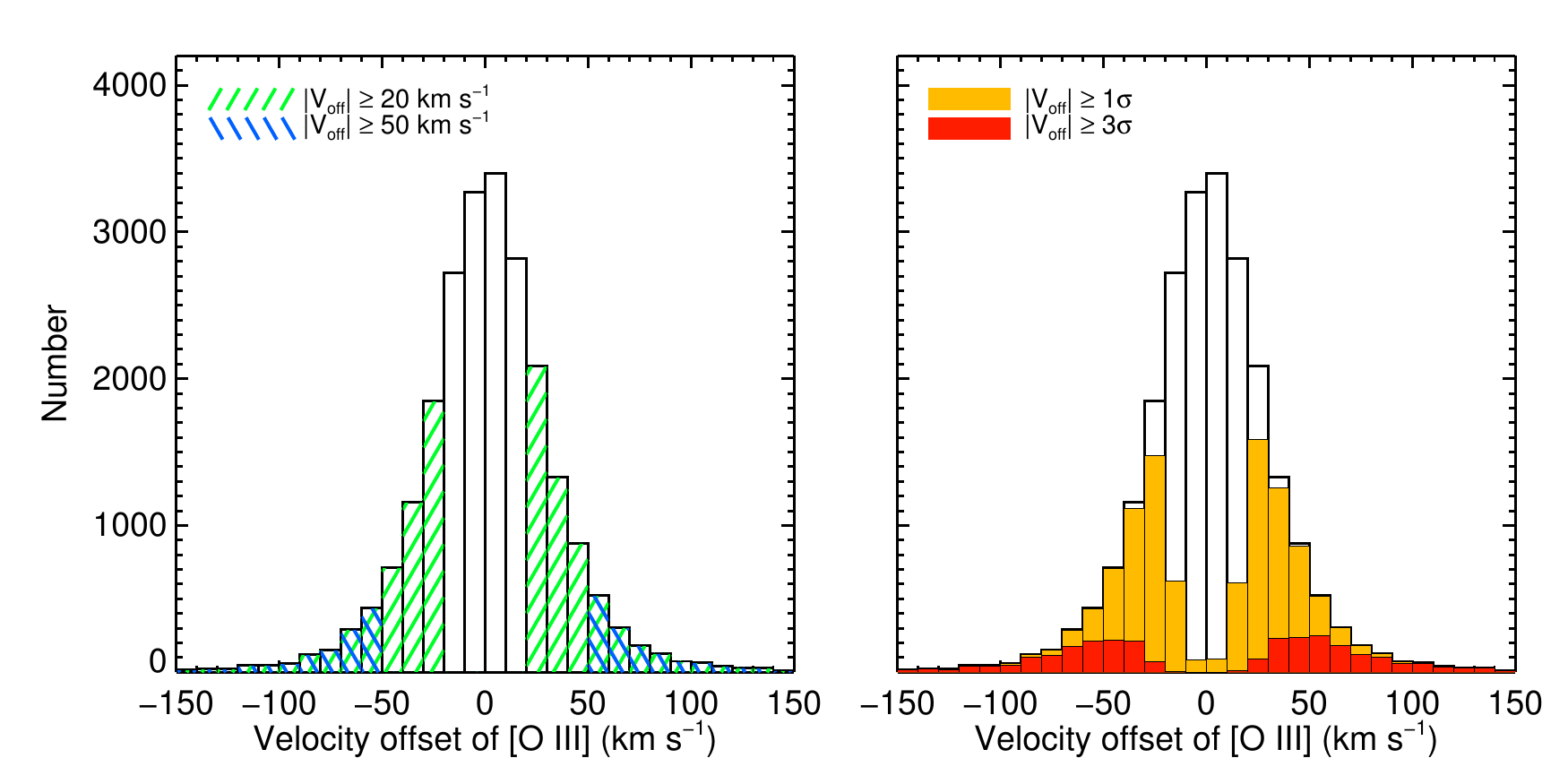}
\caption{Distributions of the [\OIII] velocity offsets with respect to the systemic velocity for $\sim$23,000 SDSS type 2 AGNs. 
The green- and blue-colored areas include AGNs with the velocity offset
$\geq$ 20 \kms\ and 50 \kms, respectively (left panel) while the orange- and red-colored areas include AGNs with the velocity offset larger than the measurement
error (1-$\sigma$ and 3-$\sigma$, respectively; right panel). }
\end{figure*}

\begin{figure*}
\centering
\includegraphics[width=0.8\textwidth]{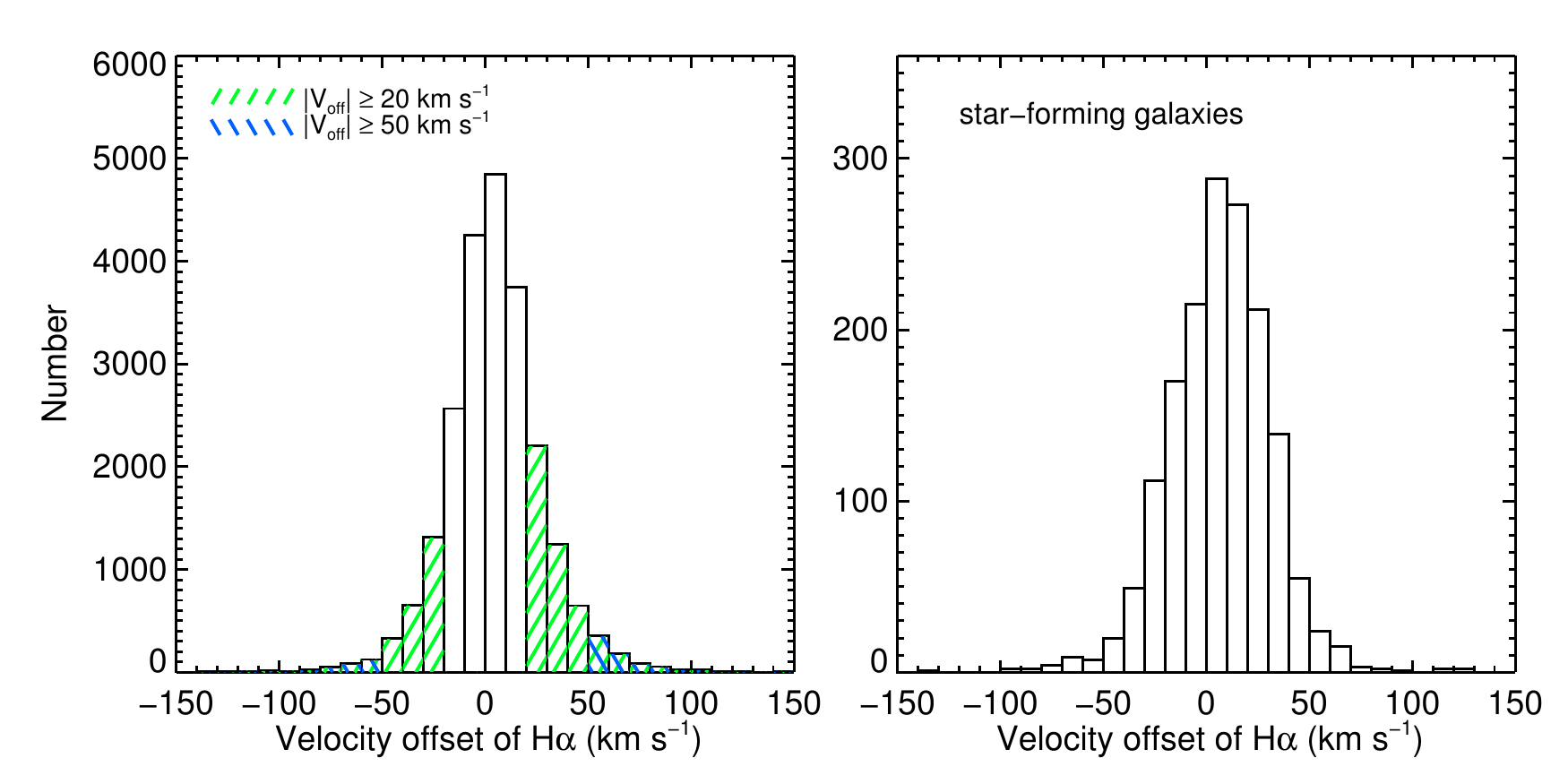}
\caption{Distributions of the \Ha\ velocity offsets with respect to the systemic velocity for the AGN sample (left panel) and the star-forming galaxies (right panel). 
The green- and blue-colored areas include AGNs with the velocity offset $\geq$ 20 \kms\ and 50 \kms, respectively.} 
\end{figure*}

\section{Sample Selection \& Analysis}
\subsection{Sample selection}
We utilized the SDSS DR 7 by matching the KIAS value-added galaxy catalog \citep{ch10} and the MPA-JHU value-added catalog\footnote[4]{\url{http://www.mpa-garching.mpg.de/SDSS/}} for the spectroscopic information. We selected galaxies at redshift range $0.02 < z < 0.1$ with signal-to-noise ratio (S/N) $\ge$ 10 for the continuum, and S/N $\ge$ 3 for the four emission lines, H$\beta$, [\OIII] $\lambda$5007, \Ha, and [\NII] $\lambda$6584.
Then the flux ratios of these four emission lines were used to classify AGNs \citep{ba81}. 
By adopting the demarcation line for AGNs and star-forming galaxies suggested by \citet{ka03}, we selected 60,018 type 2 AGN-host galaxies from the SDSS DR 7 catalog. As a comparison sample, we also selected $\sim$1,600 star-forming galaxies with an additional constraint, log([\NII]/\Ha) $<$ 0.7, which was implemented to avoid a possible contamination of weak AGNs.

To ensure robust results on the AGN outflow statistics, we further restricted the sample by selecting AGNs that have reasonably small measurement uncertainty of the velocity offset. In general the radial velocity errors depend on the amplitude (peak)-to-noise ratio (A/N) 
of the targeted emission lines \citep{sa06}. 
Using the amplitude of the emission lines and the noise of the starlight-subtracted continuum at the 5030\AA\ to 5130\AA\ range, we calculated the A/N ratio for the [\OIII] and \Ha.
Then we selected 24,356 AGNs with the A/N ratio $\geq$ 5 for both [\OIII] and \Ha, which is $\sim$41\% of the type 2 AGNs initially
selected from the SDSS DR 7. In other words, we removed AGNs with weak emission lines,
for which the measurement of the emission line velocity offsets would not be reliable.

\subsection{Velocity and Velocity Dispersion Measurements}
We measured the velocity offset of the emission lines from the starlight-subtracted spectrum. To subtract the stellar absorption lines, we built a best-fit stellar template with the penalized pixel-fitting code \citep[pPXF,][]{ce04}, using 47 MILES simple stellar population models with solar metallicity and age spanning from 60 Myrs to 12.6 Gyrs \citep{sb06}. In this procedure we measured the velocity of the luminosity-weighted stellar component of the host galaxy (redshift), and adopted it as the systemic velocity. After constructing 100 mock galaxy spectra by 
randomizing the noise, we fit the stellar absorption lines for each spectrum to estimate the uncertainty. By adopting the 1-$\sigma$ dispersion of measurements as the measurement error,
we obtained the mean error of the systemic velocity as $\sim$9$\pm$5 \kms.

From the starlight-subtracted spectrum, we fit the narrow-emission lines, \Ha, H$\beta$, [\NII]-doublet, [\OI] $\lambda$6300, and [\OIII] $\lambda$5007, by utilizing MPFIT code \citep{ma09} with a single-Gaussian function. 
%{\bf We measured the forbidden-line velocities individually because it is more appropriate for AGNs with outflows \citep[cf.][]{sa06,oh11}.} 
By performing visual inspection of the line fitting results, we removed any contamination from double-peaked line AGNs ($\sim$3.5\%), type 1 AGNs with a broad \Ha\ line ($\sim$2.0\%), or low-quality spectra/fitting ($\sim$0.5\%), finalizing 22,906 AGNs as the final AGN sample. 
We find that $\sim$11\% AGNs in the final sample show the presence of a broad wing in the [\OIII] emission line profile. To improve the fitting results, we fit [\OIII] with a double-Gaussian function 
for these AGNs, and adopted the peak wavelength of the total profile as the representative 
center of the [\OIII] line.
Finally, we calculated the velocity offset of each emission line by measuring the shift of the line peak with respect to the systemic velocity.

In addition to the velocity offset, we measured the line dispersion by calculating the second moment of the continuum-subtracted emission line profile:
\begin{eqnarray}
\sigma_{line}^2 (\lambda) = {\int \lambda^2 f(\lambda) d\lambda \over \int f(\lambda) d\lambda} - \lambda_0^2 
\end{eqnarray}
where $f(\lambda)$ is the flux at each wavelength, and $\lambda_0$ is the first moment of the line. 
%\citep{pe04}. 
The measured line dispersions were corrected for the wavelength-dependent instrumental resolution of the SDSS spectroscopy, then converted to velocity.

To obtain the measurement error of the velocity and the line dispersion for each object,
we constructed 100 mock spectra by randomizing the noise. By fitting the emission lines in 
each mock spectrum, we obtained the 1-$\sigma$ dispersion of the measurements as the 
measurement error. For the AGN sample, the mean measurement error of the velocity is $\sim$16$\pm$10 \kms\ and $\sim$7$\pm$7 \kms, respectively for [\OIII] and \Ha. 
In the case of the line dispersion, the mean measurement error is $\sim$23$\pm$35 \kms\ and $\sim$10$\pm$10 \kms, respectively for [\OIII] and \Ha.
By adding the errors of the emission line velocity and the systemic velocity in quadrature, 
we determined the total error of the velocity offset as $\sim$19$\pm$10 \kms\ and $\sim$12$\pm$7 \kms, respectively for the [\OIII] and \Ha\ lines.

By performing the same error analysis for the star-forming galaxy sample, 
we obtained the systematic velocity error as $\sim$17$\pm$8 \kms, which is larger than
the case of the AGN sample due to the higher uncertainty of the stellar population modeling.
In the case of emission lines, we estimated the velocity error $\sim$4$\pm$3 \kms\ and 
$\sim$1$\pm$1 \kms, respectively for [\OIII] and \Ha, and the line dispersion 
error $\sim$22$\pm$16 \kms\ and $\sim$9$\pm$10 \kms, respectively for [\OIII] and \Ha. 

\begin{figure*}
\centering
\includegraphics[width=0.95\textwidth]{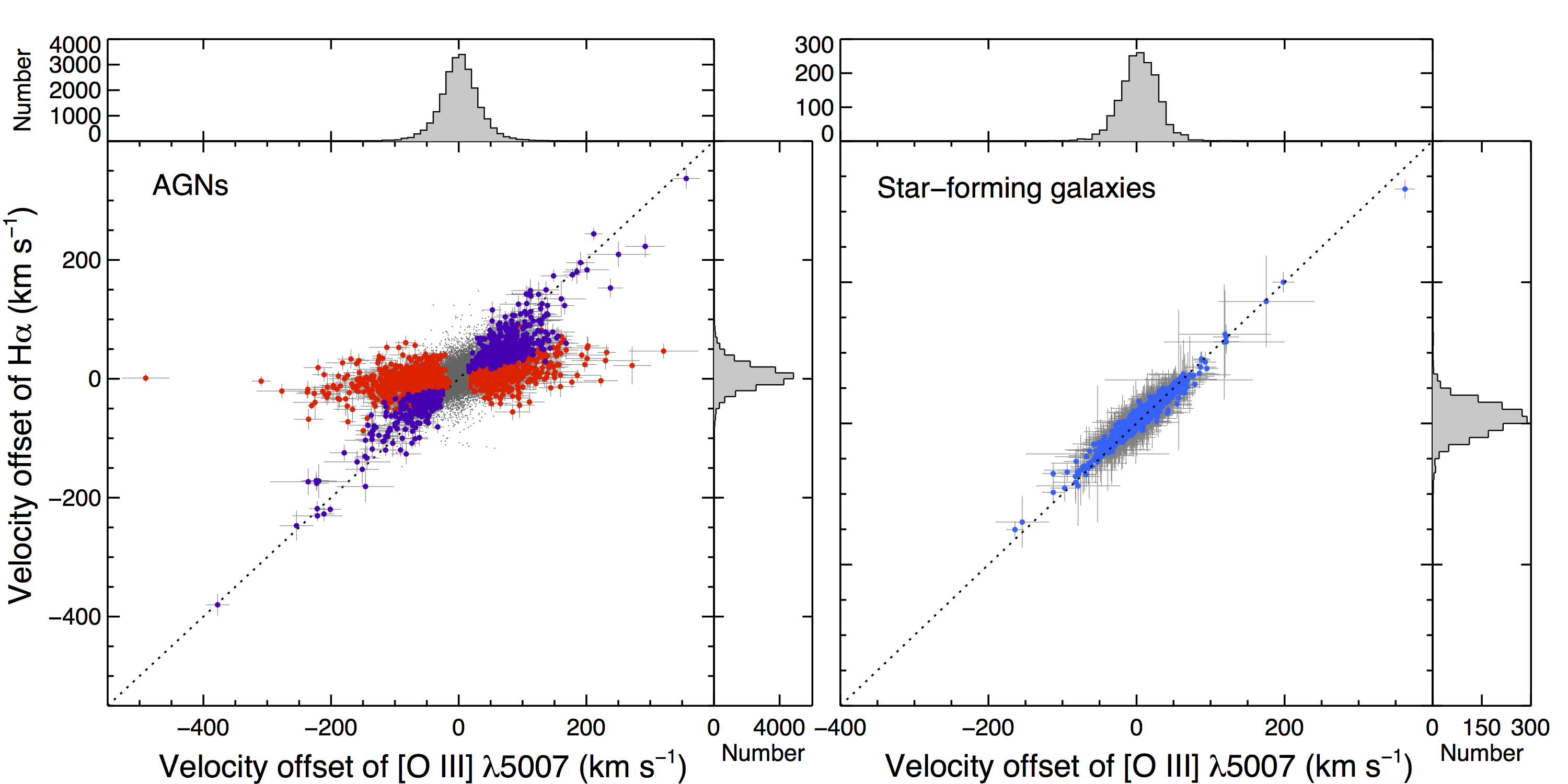}
\caption{Comparison of the velocity offsets of [\OIII] and \Ha\ for AGNs (left) and
star-forming galaxies (right). AGNs with the [\OIII] velocity offset detected 
with 3-$\sigma$ limit are denoted with colored symbols while AGNs with no or
weak ($<$ 3-$\sigma$) [\OIII] velocity offset are represented by gray points.
AGNs in Group A (red points) are defined i.e., the velocity difference between
[\OIII] and \Ha\ is larger than 3-$\sigma$ uncertainty, 
while AGNs in Group B (purple points)
are defined with comparable velocities between [\OIII] and \Ha\ within 3-$\sigma$
uncertainty. The one-to-one correspondence is denoted with the dotted lines.} 
\end{figure*}

\section{Results}

\subsection{Velocity offset of the narrow emission-lines}

In Figure 1 (left), we present the distribution of the line-of-sight velocity offsets of [\OIII] with respect to the systemic velocity. The distribution shows a peak near 0 \kms\ with a dispersion $\sim$36 \kms. 
If we restrict the sample with the [\OIII] velocity offset $\geq$ 20 \kms\ by considering the error of the velocity offset measurement (see Section 2.2), then 47\% of AGNs show velocity offset. If we further restrict the sample with the [\OIII] velocity offset $\geq$ 50 \kms, as often adopted for selecting outflows in the previous studies of type 1 AGNs \citep{bo05,ko08,zh11}, the fraction decreases to $\sim$12\%. 
The lower fraction of the [\OIII] velocity offset in type 2 AGNs is expected 
from the unification model since the direction of the outflows is close to the plane 
of the sky. 
By considering the measurement uncertainty of the velocity offset for individual objects, 
we can also restrict the sample. In this case, the fraction of AGNs with the
[\OIII] velocity offset with 3-$\sigma$ and 1-$\sigma$ detection is 
$\sim$12\% and $\sim$48\%, respectively (Figure 1, right). 
We will further discuss the outflow statistics in Section 4.3.

While the [\OIII] emission line has been mainly used as an indicator of AGN outflows, the Balmer lines also show velocity offsets with respect to the systemic velocity. Figure 2 (left) presents the distribution of the line-of-sight velocity offsets of \Ha, 
which shows a peak at $\sim$5 \kms\ with a dispersion $\sim$25 \kms. 
The broader distribution of the [\OIII] velocity offsets than that of 
the \Ha\ velocity offsets may indicate that \Ha\ is less affected 
by AGN activity than the [\OIII] line. 
In the case of the \Ha\ offset fraction, $\sim$33\% ($\sim$5\%) of AGNs show the velocity offset $\geq$ 20 (50) \kms, which is smaller than the case of [\OIII].

We note that, however, the 3\arcsec\ SDSS fiber size is larger (1 -- 6 kpc depending on the redshift of object) than the typical size of the NLR ($<$ $\sim$1 kpc). Hence, the spatially integrated spectra may also include star-forming regions outside of the active nuclei, although the flux-weighted line flux ratios indicate that the emission mainly originate from AGN. If this is the case, \Ha\ is more severely affected than [\OIII] by the contamination. 
For star-forming galaxies, the distribution of \Ha\ velocity offset 
shows a dispersion of $\sim$25 \kms, which is comparable to that of the AGN sample, 
supporting that the velocity of \Ha\ is possibly contaminated by the star-forming region in the gas disk.

\begin{figure*}
\centering
\includegraphics[width=0.48\textwidth]{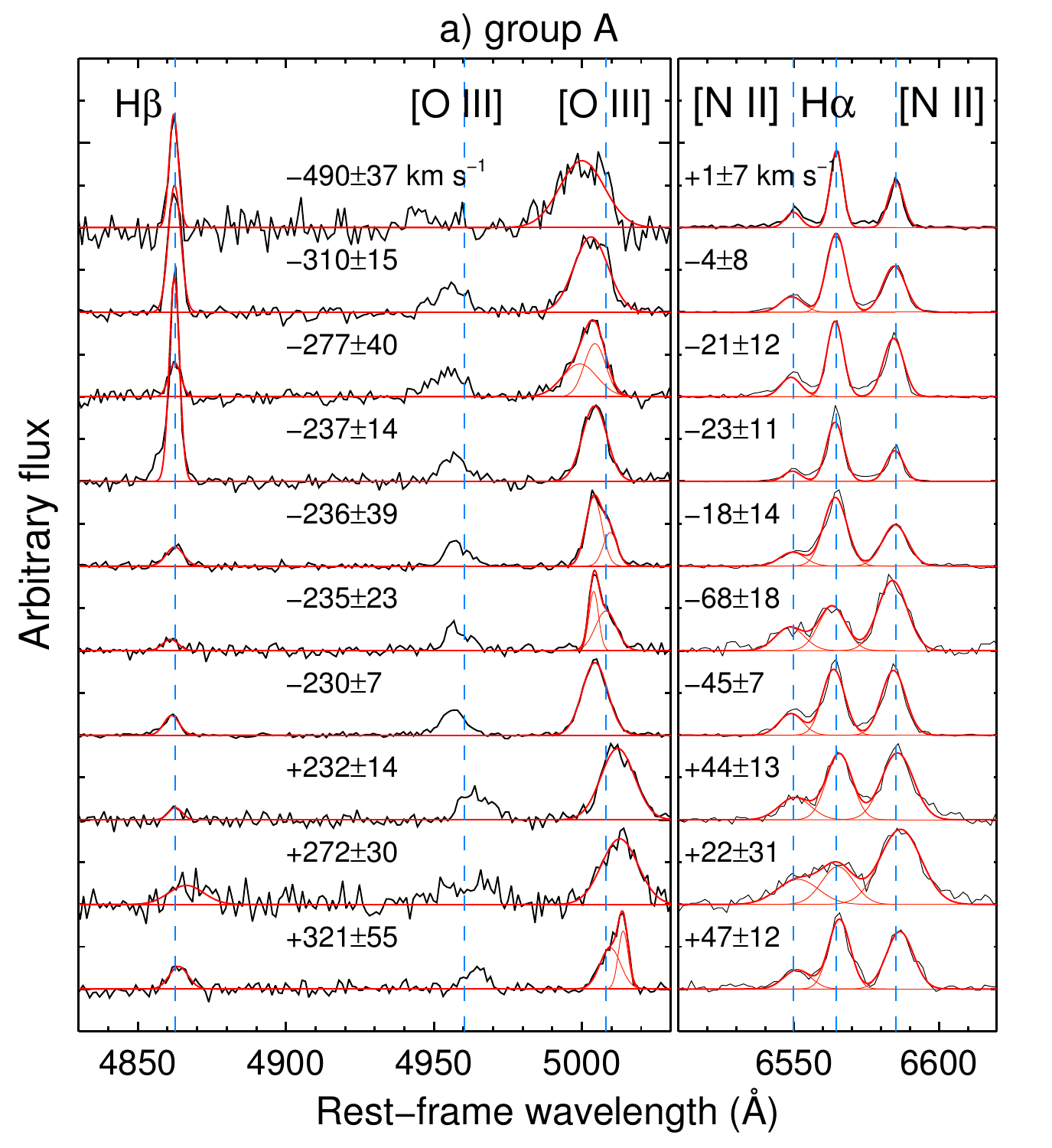}
\includegraphics[width=0.48\textwidth]{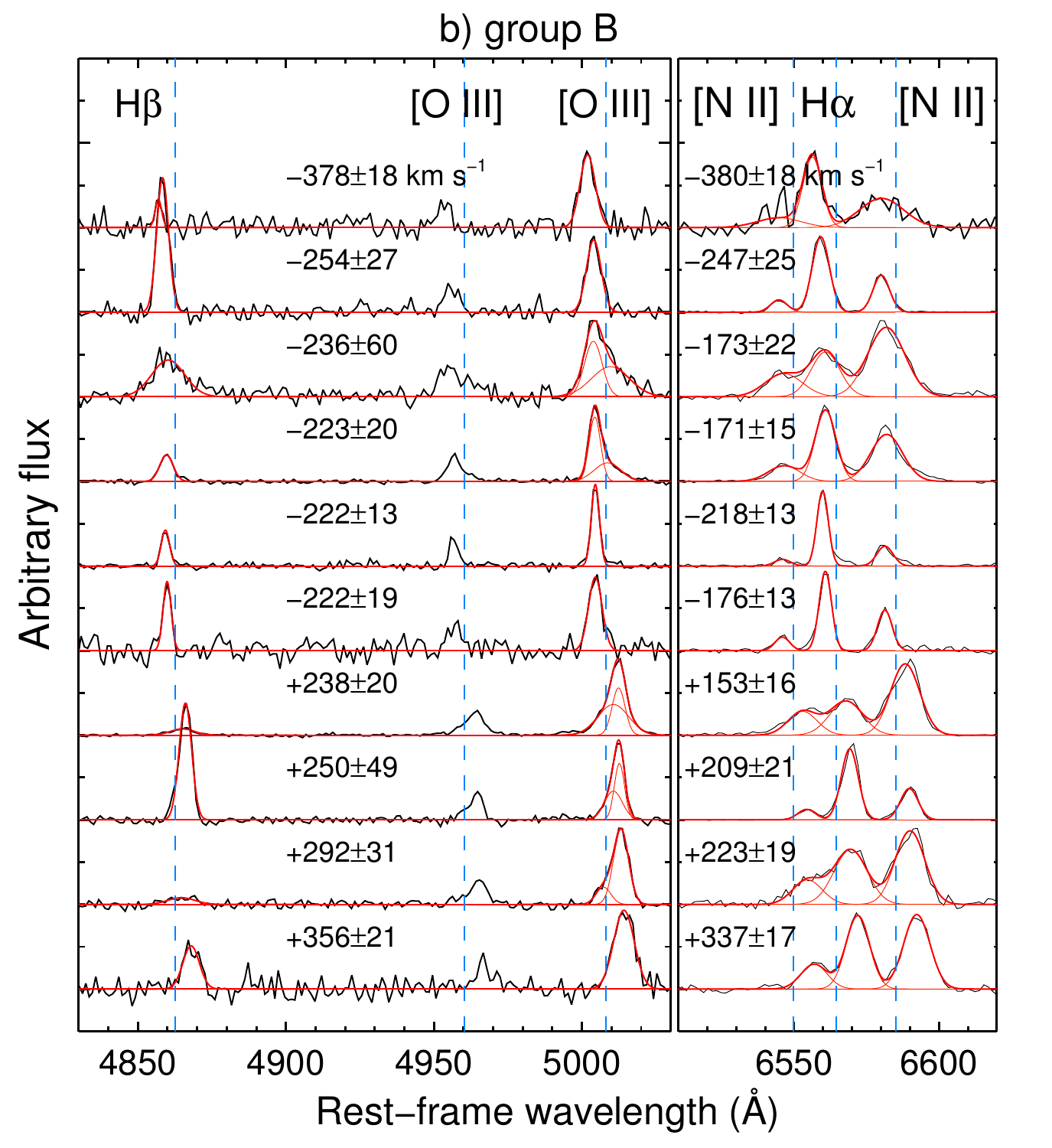}
\caption{Examples of the starlight-subtracted spectra (black solid line) of 10
AGNs with the largest [\OIII] velocity offset in a) Group A, and b) group B.
The best-fit models for the Balmer lines, the [\NII]-doublet, and the [\OIII] line 
are overlaid (red and orange lines). The expected location of each emission line center
based on the systemic velocity is indicated by blue dashed lines.
The measured velocity offsets of [\OIII] and \Ha\ are indicated in each panel.}
\end{figure*}

\setcounter{figure}{3}
\begin{figure}
\centering
\includegraphics[width=0.48\textwidth]{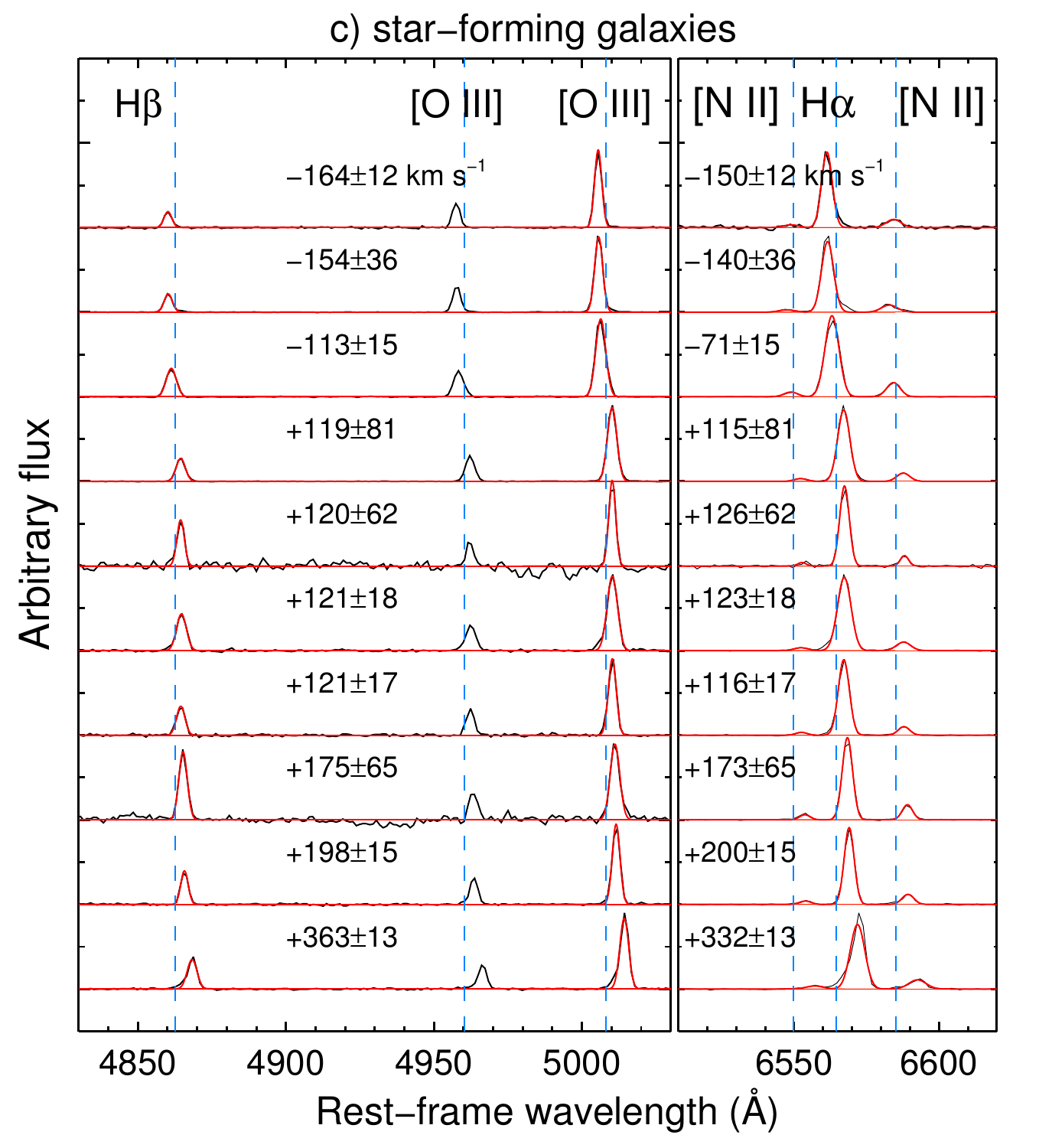}
\caption{c) Examples of the starlight-subtracted spectra of 10 star-forming galaxies with the
largest [\OIII] velocity offset.}
\end{figure}

\subsection{Comparison of velocity offset}

We compare the velocity offsets of [\OIII] and \Ha\ in Figure 3.
For the majority of AGNs, the [\OIII] velocity offset is larger than that of \Ha,
while we find some fraction of AGNs with comparable velocity offset between [\OIII] and \Ha.
Note that the velocity difference between [\OIII] and \Ha\ has been used for identifying AGN outflows in the previous studies \citep[e.g.,][]{bo05,ko08,cr10}. 
We also find that $\sim$18\% of AGNs show effectively no velocity offset in both [\OIII] and \Ha\ ($<$ 1-$\sigma$). For these AGNs, it is possible that the velocity offset
is relatively weak or that the gas motion effectively canceled out each other 
in the spatially integrated SDSS spectra. 
   
To investigate the gas kinematics of [\OIII] in detail, we identify 2,855 AGNs 
with a robust [\OIII] velocity offset detection ($>$ 3-$\sigma$) as shown with 
colored dots in Figure 3. As discussed in the previous section, the fraction of these AGNs is $\sim$12\% of the type 2 AGNs. 
For a large number of these AGNs, the velocity offset of \Ha\ is not detected 
($<$ 3-$\sigma$) or relatively weak compared to the [\OIII] velocity offset. 
Nevertheless, we find some fraction of AGNs, 
for which the \Ha\ velocity offset is detected ($>$ 3-$\sigma$) and at the same time 
[\OIII] and \Ha\ show a comparable velocity offset (i.e., the velocity difference between
[\OIII] and \Ha\ is less than 3-$\sigma$ uncertainty) as shown with purple dots in Figure 3. 
Since the latter group of AGNs show distinct gas kinematics, we will compare them with
the rest of AGNs. For this purpose, we call the AGNs with [\OIII] velocity offset larger
than \Ha\ as Group A, while we call the AGNs with comparable velocities between [\OIII] and \Ha\ as Group B.

Group A is composed of 2,061 objects ($\sim$9\%; red dots in Figure 3),
showing relatively large [\OIII] velocity offsets compared to \Ha\ velocity offsets.  
Note that there are 495 AGNs in Group A, which show opposite signs between
the [\OIII] and \Ha\ velocities. However, for most of these AGNs, the \Ha\
velocity offset is very weak ($\sim$ 10 \kms) and smaller than 3-$\sigma$ uncertainty.
Thus, it is not clear whether these AGN truly have opposite signs between
[\OIII] and \Ha\ velocities. In contrast, 28 AGNs with a robust 
\Ha\ velocity offset detection ($>$ 3-$\sigma$) show opposite signs between
[\OIII] and \Ha. It is possible that \Ha-emitting gas has dramatically different
kinematics compared to [\OIII]-emitting gas in these AGNs. However, the velocity
offset of these objects is on average very small (i.e., \Ha\ velocity offset  = 36 \kms\ with a standard deviation 10 \kms; [\OIII] velocity offset = $-$68 \kms\ with a standard deviation 37 \kms; note that all of them have redshifted \Ha\ and blueshifted [\OIII]). 
Although the origin of the opposite direction of the velocity offsets is not fully understood without spatially resolved data, it is probably 
due to the more complex kinematics of the \Ha-emitting gas, which can be
attributed by asymmetric distribution in a rotating disk, 
while the [\OIII]-emitting gas is mainly concentrated at the center.
Since these AGNs are only 1\% of Group A and the \Ha\ velocity offset is relatively 
small, we simply include them in Group A without further classification
for the following analysis. 

Group B consists of 794 AGNs ($\sim$3\%; purple dots in Figure 3), showing that
[\OIII] and \Ha\ have comparable velocities within the 3-$\sigma$ uncertainty. 
If we change the definition of Group B with 2-$\sigma$ uncertainty (i.e., the velocity difference
between [\OIII] and \Ha\ $<$ 2-$\sigma$), then the number of AGNs in Group B decreases to 727 objects
while the number of AGNs in Group A increases to 2128 objects. However, the choice
of the $\sigma$ limit does not significantly change the results in the following 
analysis.
There are AGNs with the velocity offset of \Ha\ larger than that of 
[\OIII], while the velocity difference between \Ha\ and [\OIII] is smaller 
than 3-$\sigma$ for all AGNs in Group B. 
If we change the $\sigma$ limit for the velocity difference between
[\OIII] and \Ha, only 4 and 24 objects show larger \Ha\ velocity than [\OIII]
velocity, respectively for 2-$\sigma$ and 1-$\sigma$ limit. 

In Figure 4, we present the examples of the starlight-subtracted spectra, showing 
that the [\OIII] emission line is clearly shifted from the expected wavelength based on the systemic velocity, while the Balmer and the [\NII] lines show no and similar offset, respectively for Group A and B.

In contrast, for star-forming galaxies [\OIII] and \Ha\ show comparable velocity offsets (Figure 4c) and we do not find any star-forming galaxies with significantly 
different velocities between [\OIII] and \Ha. This is dramatically different from 
the trend we described for the AGN sample.
The distributions of [\OIII] and \Ha\ velocity offsets 
are similar with a 1-$\sigma$ dispersion of $\sim$30 \kms. The galaxies with a large 
velocity offset ($\ge$ 30 \kms) present an asymmetric disk in the SDSS image, suggesting that the velocity offset 
detected in the spatially integrated spectra is presumably due to the asymmetric 
distribution of star-forming regions in the rotating disk. 
The fact that star-forming galaxies and Group B show similar \Ha\ velocity offsets 
with respect to [\OIII] may suggest that the \Ha\ velocity of Group B is also affected by 
star formation. However, star-forming galaxies and Group B do not necessarily have the same origin
of the velocity offset since the emission line flux ratios are clearly different
(see Section 2.1). Also, the dramatic difference in the distribution of the [\OIII] and \Ha\ 
line dispersions further supports that Group B may have a different origin (see Figure 5 and Section 3.3). Possible scenarios on the origin of the velocity offsets in Group B are further discussed in Section 4.2.

\begin{figure*}
\centering
\includegraphics[width=1.0\textwidth]{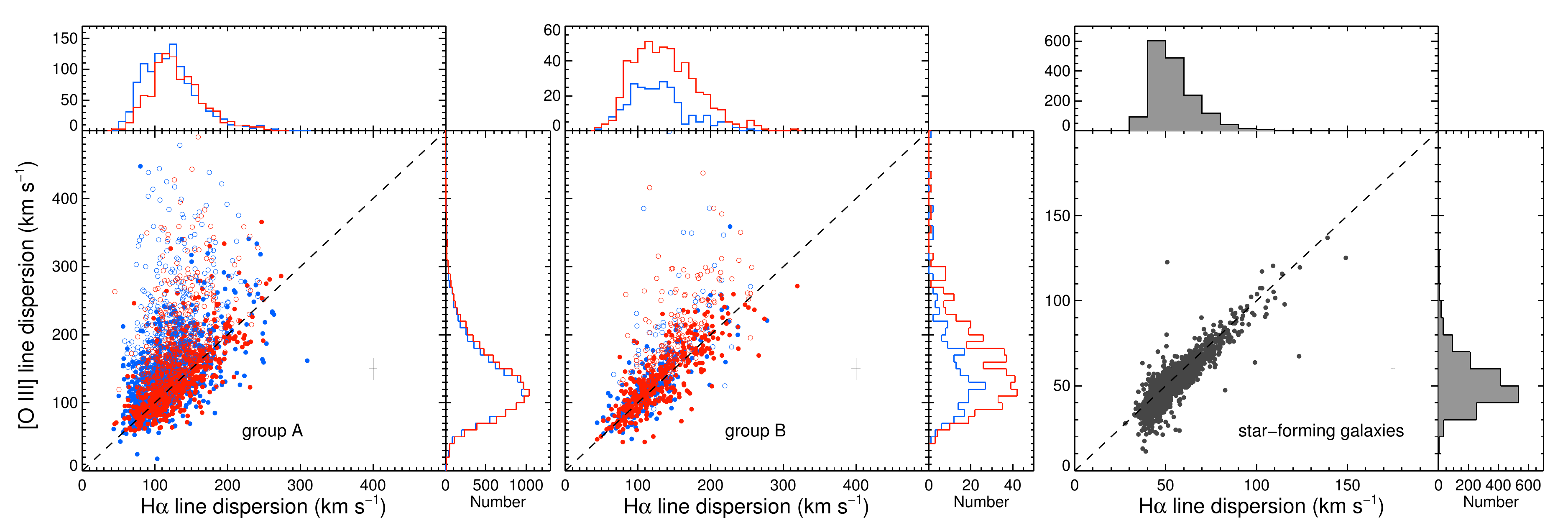}
\caption{Comparison of the line dispersions between [\OIII] and \Ha\ for Group A (left),
 Group B (middle), and the star-forming galaxies (right). 
Blue and red dots represent AGNs with the blueshifted- and redshifted [\OIII], respectively
while open (filled) circles indicate the presence (absence) of an asymmetric wing feature 
in the [\OIII] line profile. The mean errors of the line dispersion measurements are 
denoted at the bottom right.
Note that the dispersion range for the star-forming galaxies is a factor of two smaller than that of AGNs.}
\end{figure*}

\begin{figure}
\centering
\includegraphics[width=0.5\textwidth]{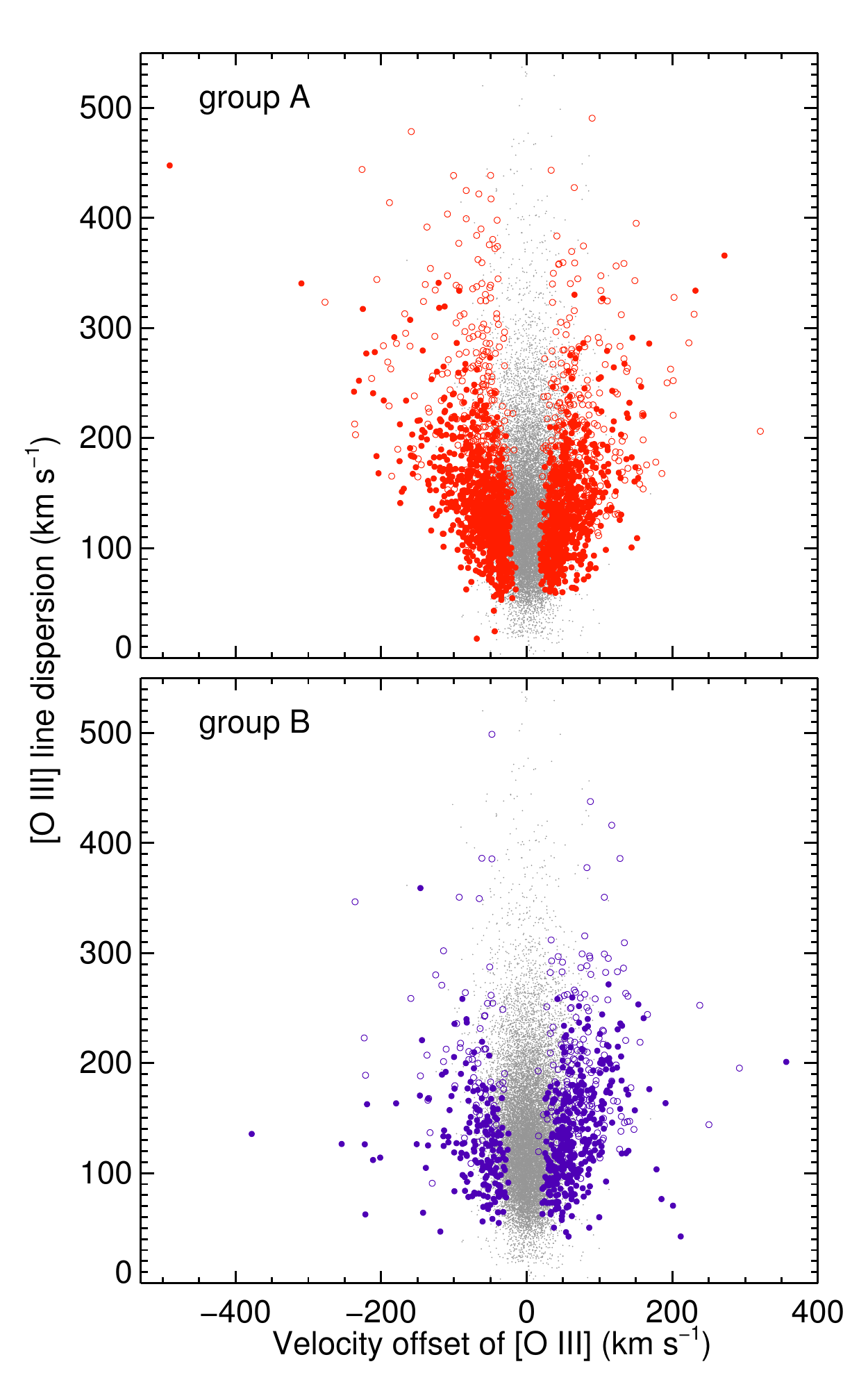}
\caption{Comparison of the velocity offset and line dispersion of [\OIII] for Group A (top) and Group B (bottom).
Symbols are the same as in Figure 3.}
\end{figure}

\subsection{Comparison of line dispersion with velocity offset}

We compare the line dispersions of [\OIII] and \Ha\ in Figure 5. 
In Group A, the [\OIII] line dispersions show a wide distribution 
(mean $\sim$161 \kms and 1-$\sigma$ dispersion $\sim$67 \kms)
with a long tail toward the high line dispersion, while H$\alpha$ shows narrower distribution (mean $\sim$126 \kms\ and 1-$\sigma$ dispersion $\sim$38 \kms),
suggesting the presence of high-velocity gas in the [\OIII]-emitting region. 
For individual AGNs, [\OIII] is on average $\sim$31\% broader than \Ha,
indicating that the outflows with a wide opening angle 
contributes to the line-of-sight velocity dispersion of [\OIII] while the \Ha\ line
is less affected. The large [\OIII] line dispersions are mostly related to the presence of a broad wing in the line profile (open circles). 
In the case of Group B, the line dispersions of [\OIII] and \Ha\ are similar to each other
except for a small number of outliers.

In contrast, the star-forming galaxies show much narrower emission lines than AGNs,
and comparable line dispersions between [\OIII] and \Ha\ (Figure 5, right panel), 
as expected from the lack of asymmetric broad wing features (see Figure 4c).
Although most of the resolution-corrected line dispersions are below the SDSS instrumental resolution, 
%$\sim$70 \kms, 
it is clear that  
[\OIII] and \Ha\ in star-forming galaxies have comparable random velocities
while [\OIII] in AGNs has a systematically larger random velocity than \Ha.  

In Figure 6, we compare the velocity offset with the line dispersion. 
In Group A, large [\OIII] velocity offsets are preferentially found among AGNs 
with large line dispersion, implying that the [\OIII] line widths are also increased due to
the AGN-driven outflows, while such a trend is weaker in Group B.        

The gravitational potential of the galaxy bulge is mainly responsible for the broadening
of the [\OIII] emission line when the outflow is not dominant as shown by the previous works in 
comparing the [\OIII] line widths with stellar velocity dispersions \citep[e.g.,][]{nw95,kx07}.
When AGN outflow is strong, however, the emission lines can be additionally broadened due to 
the gas motion in biconical outflows with a wide opening angle, 
as manifested by the presence of broad wings in the [\OIII] emission line profile.  

We note that some AGNs in Group A show large [\OIII] line dispersions ($\ge$ 300 \kms),
and relatively low velocity offsets ($<$ 100 \kms). Most of these
objects have a broad wing in the [\OIII] line profile (open circles), indicating
the presence of high velocity gas. It is possible that the outflow direction of these
objects is almost perpendicular to the line-of-sight, resulting in a small projected 
velocity offset while the line width is additionally broadened due to the radial gas motion 
in a wide opening angle, which includes blueshifted and redshifted components to the line-of-sight.

The trend between velocity offset and line dispersion can be also related to the 
dust obscuration \citep{cr10}. Dust-patches in the inner stellar disk may preferentially obscure one side of the biconical outflows,
depending on the orientation angle between the stellar disk and the outflow direction,
hence either blueshifted or redshifted velocity offset is detected in the luminosity-weighted integrated
spectroscopic observations. If there is a biconical outflow and no internal dust extinction, one may find 
a larger line width than expected from the galaxy bulge potential, 
while no significant velocity offset is expected since the approaching and receding gas 
motions will cancel out each other.

\subsection{Velocity offset vs. Eddington ratio}

In this section we compare the [\OIII] velocity offset with the Eddington ratio.
We calculate AGN bolometric luminosity using the equation suggested by \citet{ne09} for type 2 AGNs:
\begin{eqnarray}
\textrm{log}L_{\rm bol} =3.8+0.25\textrm{log}L([\OIII] \lambda 5007){} \nonumber\\
{}+ 0.75\textrm{log}L([\OI] \lambda 6300), \rm
\end{eqnarray}
where $L$([\OIII] $\lambda$5007) and $L$([\OI] $\lambda$6300) are the extinction-corrected luminosities of [\OIII] and [\OI], respectively.
We correct for extinction using the extinction law suggested by \citet{ca99}, assuming the intrinsic Balmer decrement \Ha/H$\beta$ = 2.86 \citep{ne09}. 
For this calculation, $\sim$13\% of AGNs have been excluded due to the non-detection of [\OI] emission.  
As a consistency check, we also calculate the bolometric luminosity using the equation $L_{\rm bol} = 700 \times L([\OIII] \lambda 5007)$ \citep{la09}, 
where the [\OIII] luminosity is also extinction-corrected.  
We use the stellar velocity dispersion ($\sigma_{\star}$) available from MPA-JHU catalog to infer black hole mass (M$_{\rm BH}$)  
by adopting the M$_{\rm BH}$-$\sigma_{\star}$ relation \citep{pa12,woo13}. 

We find that the Eddington ratio distributions are different between the two groups. 
In Group A, AGNs with large velocity offsets preferentially have a high Eddington ratio,
while the trend is not found in Group B (see Figure 7). 
When we use the [\OIII] luminosity as a proxy for the AGN bolometric luminosity, we obtain qualitatively the same results. 

\begin{figure*}
\centering
\includegraphics[width=0.8\textwidth]{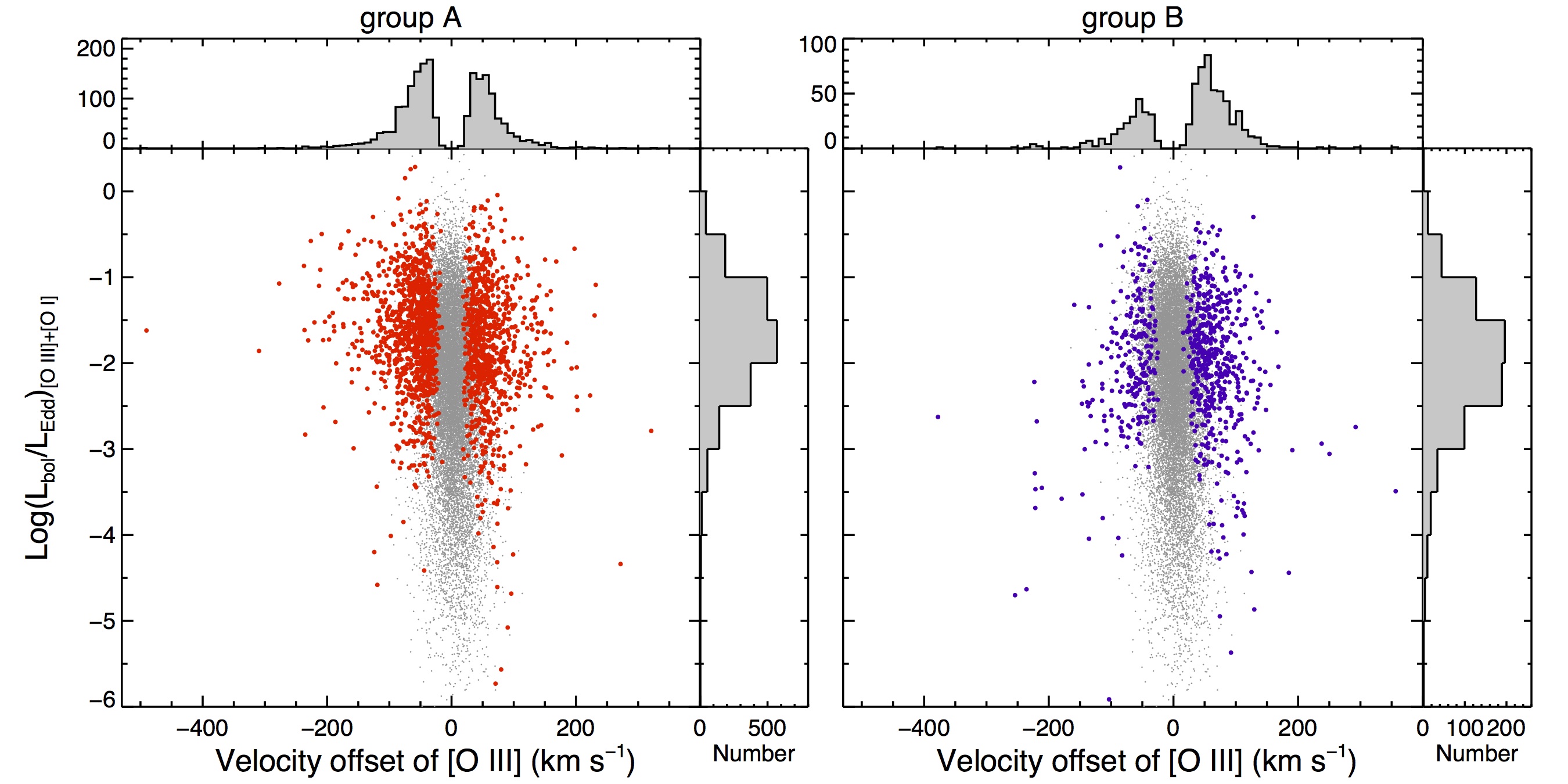}
\caption{Comparison of the [\OIII] velocity offset with the Eddington ratio for Group A (left), and Group B (right panels). 
Symbols are the same as in Figure 3.}
\end{figure*}

To examine the trend in detail, we divide the sample into five bins with different Eddington ratio range (see Table 1). 
In Group A the fraction of the detected velocity offset largely increases from 1\% to 17\% as the Eddington ratio increases,
while Group B does not show such a trend. 
Although the estimated Eddington ratios have large uncertainties, our results
are consistent with that the detected velocity offsets are due to AGN activity.

While we find that AGNs with a large [\OIII] velocity offset preferentially 
have a high Eddington ratio, we also find AGNs with a high Eddington ratio, but with
a small velocity offset. One of the possible explanations is that the low velocity offset is due to a projection effect when the outflow direction is close to the plane of the sky. 
Another possibility is that the blueshifted and redshifted components are canceled out
in the spatially integrated SDSS spectra if the dust in the stellar disk does not strongly obscure one side of the biconical outflows.

\begin{table}
\begin{center}
\caption{Number of AGNs in various Eddington ratio ranges\label{tbl-1}}
\begin{tabular}{ccccc}
\tableline\tableline
%& & \multicolumn{3}{c}{[\OIII] offset $\geq$ 20 \kms} & \multicolumn{3}{c}{[\OIII] offset $\geq$ 50 \kms} \\
%\cline{3-8}
ER & Total & A+B & Group A & Group B \\ %& Group A+B & Group A & Group B\\
(1)&(2)&(3)&(4)&(5)\\%&(6)&(7)&(8)\\
\tableline
\\
 $>$ 1                & 1350 & 293 (22\%) & 236  (17\%)& 57  (4\%) \\
\\
0.1 -- 1  & 7897 & 1391 (18\%)& 1068 (14\%)& 323 (4\%)  \\
\\
0.01 -- 0.1  & 7060 & 806 (11\%)& 518 (7\%)& 288 (4\%) \\
\\
0.001 -- 0.01  & 2550 & 120 (5\%)& 68 (3\%)& 52 (2\%) \\
\\
$<$ 0.001 & 1062 & 29 (3\%) & 13 (1\%)  & 16 (2\%) \\
\\
\tableline
\end{tabular}
\tablecomments{(1) Range of the Eddington ratios; (2) Number of the total AGNs within the Eddington ratio range; (3) Number (fraction) of both group A and B within the Eddington ratio range (4) Number (fraction) of the group A within the Eddington ratio range; (5) Number (fraction) of the group B with in the Eddington ratio range.}
\end{center}
\end{table}

\subsection{Inclination of the host galaxies}

The detected [\OIII] velocity offset can be explained with a combined model 
of biconical outflows and dust extinction in the host galaxy disk \citep{cr10}.
Since dust-patches in the disk can preferentially obscure one side of the 
biconical outflows, depending on the orientation angle between the direction of 
outflows and the disk plane, either blueshifted or redshifted [\OIII] will be
detected in the spatially integrated spectra. 

As a consistency check, we investigate the distribution of the host galaxy
inclination, respectively for the AGNs with blueshifted and redshifted [\OIII].
If the outflows are biconical, we expect that more AGNs with blueshifted [\OIII]
will be detected in face-on late-type galaxies since the redshifted component 
can be obscured by the dust in the disk. 
First, we choose late-type galaxies ($\sim$40\%) among the AGN sample, 
based on the visual morphology classification from \citet{li11}.
Second, we measure the seeing-corrected isophotal minor-to-major axis ratio b/a
on the SDSS i-band image.
For comparison, we also obtain the distribution of the axis ratio for all late-type AGN host galaxies 
in the SDSS DR 7.
 
By applying a Gaussian kernel, we present the normalized probability 
distributions in Figure 8. 
In Group A, the distribution is clearly different between the AGNs with
redshifted [\OIII]  and the AGNs with blueshifted [\OIII], particularly when b/a is larger than 0.5,
indicating that face-on galaxies are more likely to host AGNs with blueshifted [\OIII]. 
While the distribution of the AGNs with blueshifted [\OIII] is similar to
that of the general population of late-type AGN host galaxies,
the relative lack of the AGNs with redshifted [\OIII] 
is consistent with the combined model of the bipolar outflow and dust obscuration.
Since the angle between the disk plane and the outflow direction is random, 
galaxies with intermediate inclinations can obscure either blueshifted or redshifted 
[\OIII] with similar probability. As expected, no clear difference is found at intermediate b/a.
In the case of Group B, on the contrary, AGN with blueshifted {\it and} 
redshifted [\OIII] show a lack of face-on host galaxies, compared to the
general population of late-type AGN host galaxies. This may suggest that Group B galaxies
are not representative of general AGN host galaxy population (see discussion in Section 4.2).

\begin{figure*}
\centering
\includegraphics[width=1.0\textwidth]{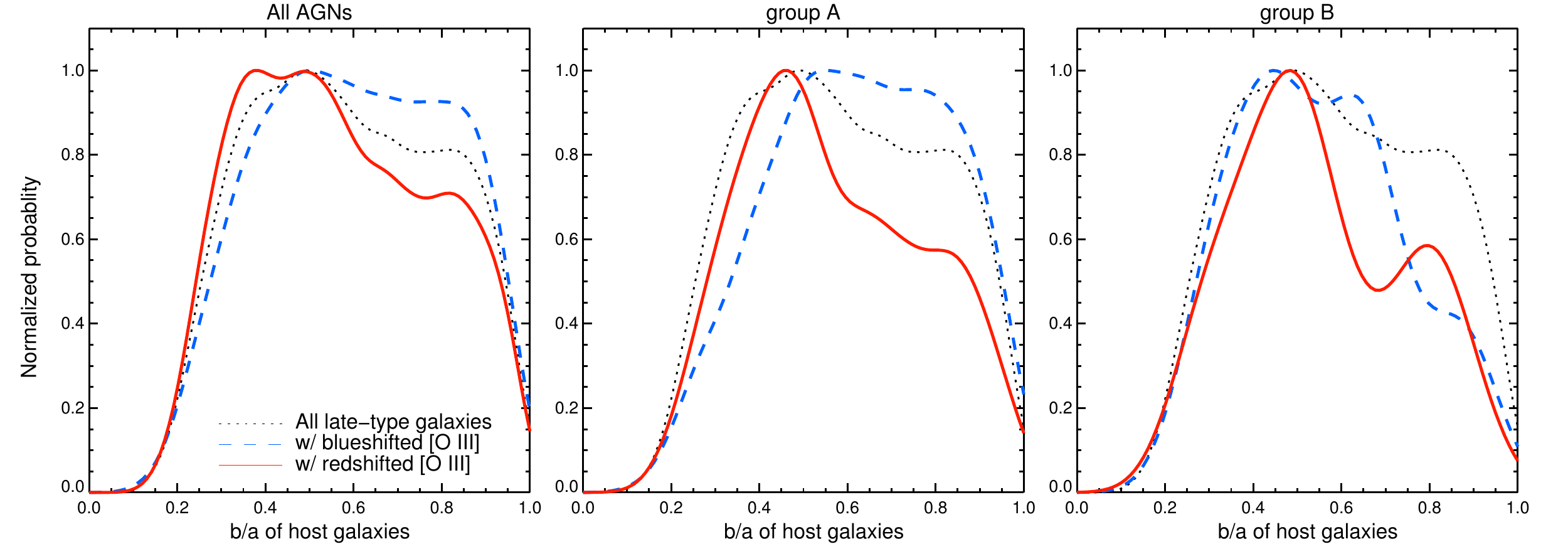}
\caption{Normalized probability distributions of the axis ratio (b/a) respectively for AGNs  with the blueshifted 
(blue dashed line) and redshifted [\OIII] (red solid line). The distributions are shown for the total sample (left), Group A (middle), and Group B (right).
The black dotted line represents all late-type AGN host galaxies from SDSS DR 7.}
\end{figure*}

\subsection{Radio properties}

We investigate whether the [\OIII] velocity offsets are related to radio jets using the FIRST survey catalog \citep{wh97}. Similarly in both Group A and Group B, $\sim$25\% of AGNs have a radio counterpart. Among the detected sources, only $\sim$1\% of AGNs show clear jet-like or elongated radio 
morphology, while the others look more or less compact (with a beam size $\sim$5\arcsec). 
The radio detection rate increases with the increasing [\OIII] velocity 
offset. In Group A, the detection rate is 17\%, 26\%, and 40\% respectively for
AGNs with the [\OIII] velocity offset smaller than 50 \kms, between 50 and 100 \kms, and larger than 100 \kms. 
A similar trend is also found in Group B. The results imply that radio jets are related with the ionized gas outflows at least for the radio-detected AGNs.

%\begin{figure*}
%\centering
%\includegraphics[width=0.95\textwidth]{fig9a.eps}
%\bigskip
%\includegraphics[width=0.95\textwidth]{fig9b.eps}
%\caption{SDSS gri-composite images of the galaxies with the [\OIII] velocity $>$ 200 \kms: 
%24 AGNs in Group A (top) and 15 AGNs in Group B (bottom).
%The [\OIII] velocity offset is labeled in each  $40\arcsec \times 40\arcsec$ size panel.
%}
%\end{figure*}

\begin{figure*}
\centering
\includegraphics[width=0.95\textwidth]{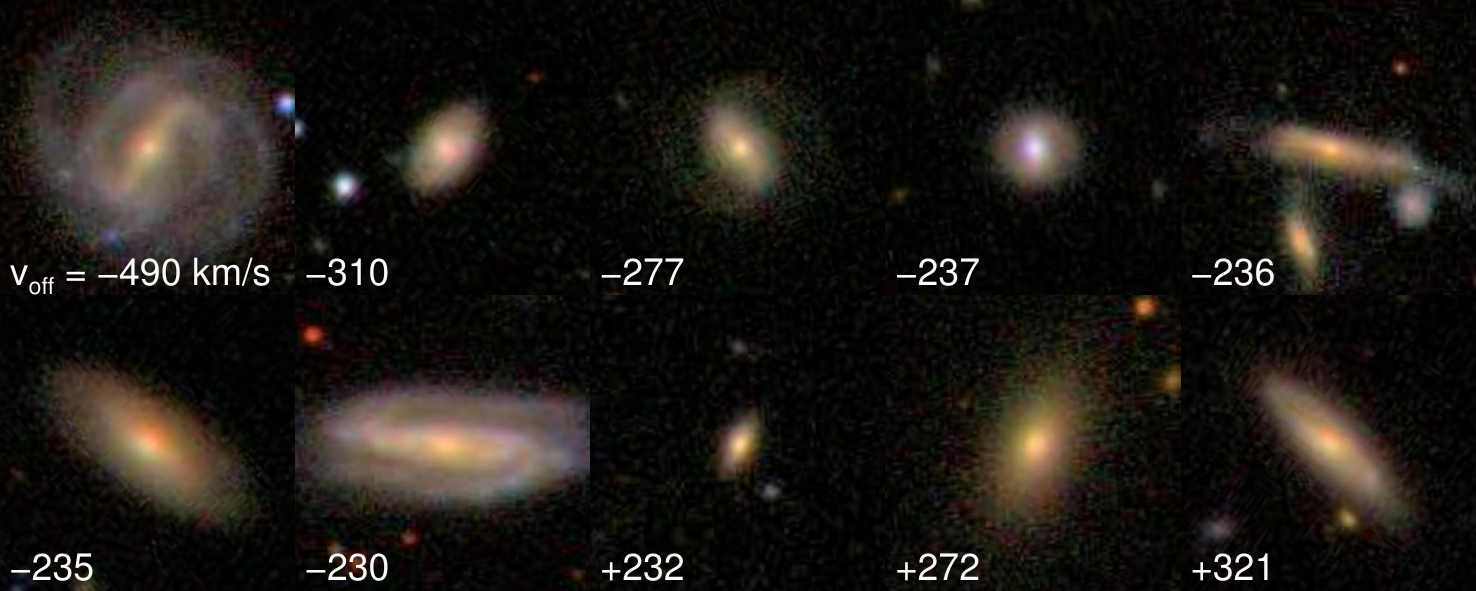} \\[3pt]
\includegraphics[width=0.95\textwidth]{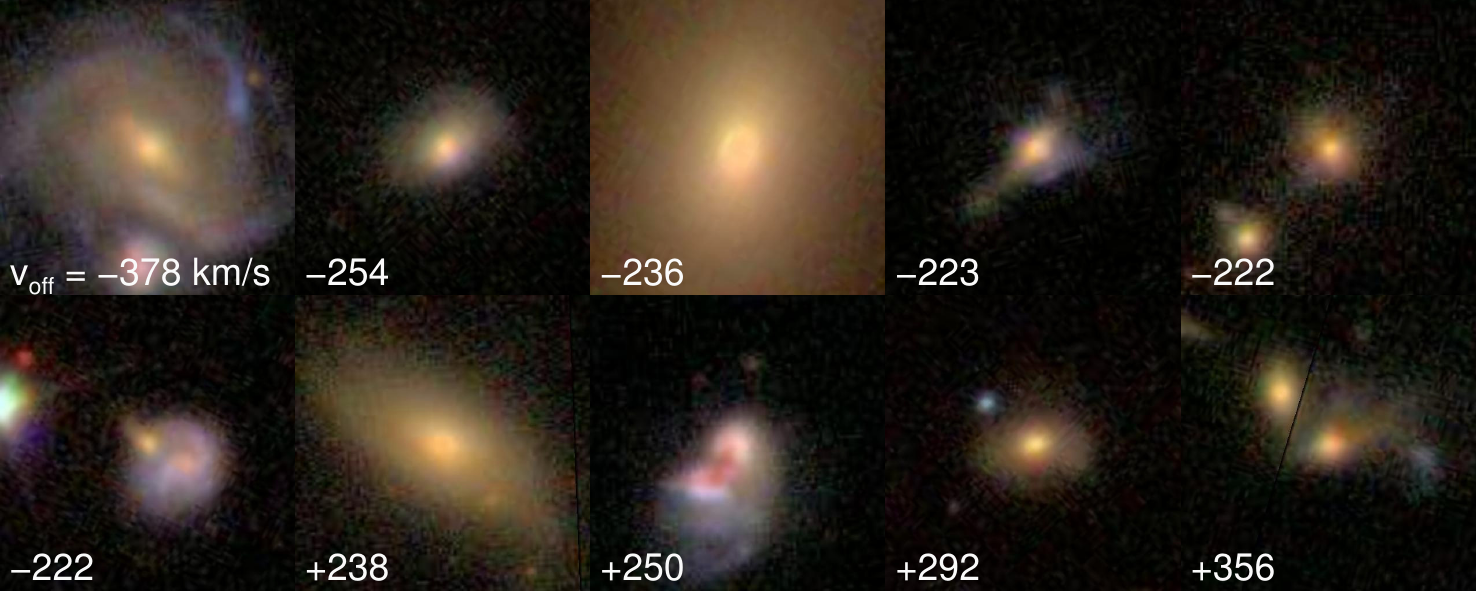}
\caption{SDSS gri-composite images of the galaxies with the largest [\OIII] velocity offset: 
10 AGNs in Group A (top) and 10 AGNs in Group B (bottom).
The [\OIII] velocity offset is labeled in each  $40\arcsec \times 40\arcsec$ size panel.
}
\end{figure*}

\section{Discussion}

\subsection{Starburst-driven vs. AGN-driven Outflows}

Although the photoionizing source of the narrow emission lines is likely to be AGNs
for given the emission line flux ratios used for our sample selection,
the nature of the velocity offsets manifested by these lines
are not straightforward to interpret. 
For instance, since AGNs and nuclear star formation activities are often coupled 
\citep[e.g.,][]{ns09,ne09,woo12}, the observed velocity offsets 
can be due to the outflows from the nuclear starburst.

One of the main differences between AGN- and starburst-driven outflows is the geometry 
of the outflows; while the AGN-driven outflows are randomly oriented with respect to the 
host galaxy, the direction of the starburst-driven outflows is in general perpendicular 
to the stellar disk \citep[][and references therein]{ve05}. 
Thus, if the detected velocity offsets are attributed to the starburst, 
the majority of spiral galaxies with intermediate to face-on inclinations (i.e., b/a $\geq$ 0.6)
are expected to show blueshifted [\OIII]. 
On the contrary, we find that $\sim$38\% and $\sim$68\% of the late-type galaxies with b/a $>$ 0.6, respectively in Group A and Group B, show 
redshifted [\OIII], suggesting that the detected velocity offsets are not 
starburst-driven.
Moreover, in the case of starburst-driven outflows, face-on galaxies
should have larger line-of-sight velocities than 
edge-on galaxies since the outflow direction is closer to the line-of-sight.
On the contrary, we find that the mean [\OIII] velocity is smaller by 
$\sim$10\% and $\sim$20\% in more inclined galaxies (b/a $\ge$ 0.6) 
than in more face-one galaxies (b/a $<$ 0.6), respectively for Group A and Group B. 
Thus, we conclude that the velocity offset of [\OIII] is not likely due to the nuclear starburst.

\subsection{Nature of the velocity offset in the group B}

As presented in Section 3.2, for a small fraction of AGNs, 
[\OIII] and \Ha\ show comparable velocity offsets with respect to the systemic
velocity. The [\OIII] velocity offsets can be interpreted as AGN-driven, 
since the emission line flux ratios are consistent with the AGN photoinoization. 
However, the origin of the comparable velocity offset of \Ha\ is unclear.
One possibility is that the deceleration of the outflows is much weaker and/or 
gas outflows are significant in the outer NLR so that \Ha\ and other low-ionization 
lines have comparable velocity offsets as [\OIII]. While the deceleration is 
expected if the outflows interact with the surrounding ISM \citep{dy10}, 
the deceleration may not be strong  in certain physical conditions, such as with a 
lower ISM density, or with more energetic outflows. 

We note that the host galaxies of Group B tend to have disturbed morphology and 
merging features, particularly for the AGNs with a large [\OIII] velocity offset; 
for example, $\sim$54\% of AGNs with the [\OIII] velocity offset $\geq$ 100 \kms, 
and all AGNs with the [\OIII] velocity offset $\geq$ 200 \kms\ in Group B present 
signatures of on-going or recent merger (e.g., see Figure 9), indicating that these galaxies
are going through a dramatic dynamical process, presumably accompanied by a strong starburst activity. 
In such a case, it is possible that the non-virialized stars and gas are kinematically 
decoupled, resulting in the velocity offset between stellar absorption lines
and the gas emission lines. 

Another possibility is that the velocity offset is caused by a binary black hole system, 
consisting of an inactive black hole and an active black hole, which is either inspiralling \citep{mm01} or recoiling \citep{ca07},
and displaced from the dynamical center of the host galaxy.
In the latter case, however, theoretical studies showed that the black hole could not 
have a large enough radius to carry the NLR after the kick \citep{me06}. 
Instead, the inspiralling active black hole with an accompanied NLR may be a viable 
explanation for the velocity offset \citep[e.g.,][]{co09,cg14}. 
Recently, \citet{cg14} find $\sim$400 AGNs in SDSS showing comparable velocity offsets between the forbidden lines and the Balmer lines, which are similar to the AGNs in Group B,
claiming that the AGNs are candidates for kinematically offset SMBH.

Using the N-body simulations, \citet{bl13} showed that during the galaxy-galaxy merging, spectroscopic observations may detect only one NLR with a significant (a few hundred \kms) velocity offset, depending on physical properties, i.e., the galaxy mass ratio, gas content, and the viewing angle. 
If the velocity offset detected in our analysis is due to the binary motions, 
the projected distance between the two black holes could be a few kpc \citep{co09,sh11}.
Considering the 3$\arcsec$ aperture size of the SDSS spectroscopy, which covers several kpc of the targets,
it is possible that the displaced active black hole and its NLR are observed through
the SDSS fiber. 
In this scenario, it is difficult to detect two active black holes since only one black hole is active for the majority of the 
merging time. Instead, the presence of double stellar cores 
may provide circumstantial evidence \citep{sh11}.
We investigated the morphology of the 142 AGNs with the 
largest ($>$ 100 \kms) velocity offsets in Group B, using the SDSS gri-composite images. We find double stellar cores only in $\sim$5\% of the AGNs, suggesting that 
%the distance between the binary black hole is already far smaller than $\sim$kpc in many of the AGNs, or 
the displaced black hole scenario is not well supported for Group B.

\subsection{AGN Outflow Statistics}

In this study, we find that $\sim$47\% of $\sim$23,000 type 2 AGNs
have an [\OIII] velocity offset larger than 20 \kms\
with respect to the systemic velocity. 
%If we divide the AGN sample into Seyfert galaxies and LINERs using 
%the criterion log([\OIII]/H$\beta$) = 0.5 \citep{vo87}, 
%the velocity offset fraction is $\sim$52\% and $\sim$45\% 
%respectively for Seyferts and LINERs. 
If we interpret the velocity offset as due to AGN outflows, the AGN outflow fraction 
based on the detected velocity offset is a lower limit because
of a couple of limitations. First, we only count the outflows when the velocity offset 
is larger than 20 \kms\ by accounting for the uncertainties of the velocity offset measurement.
Second, the measured velocity offset is the line-of-sight velocity. Thus, if the 
outflow direction is very close to the plane of the sky, we do not detect the offset
in the line-of-sight. Third, if the blueshifted and redshifted cones are symmetric 
and canceled out each other in the spatially integrated spectra, 
we do not detect the velocity offsets with respect to systemic velocity.

With these limitations, we compare the fraction of the AGN-driven outflows 
between type 1 and type 2 AGNs by combining the results from the literature. 
Various previous studies reported that $\sim$50\% of type 1 AGNs show outflows
although the analysis methods were not uniform.
For example, \citet{bo05} reported that approximately a half of 
$\sim$400 SDSS QSOs has blueshifted [\OIII] while \citet{ko08} presented that
51\% (28 out of 55) of narrow-line Seyfert 1 galaxies have the [\OIII] velocity offset 
larger than 50 \kms. A more recent study by \citet{zh11} reported that 69\% of 
$\sim$400 Seyfert 1 galaxies show the [\OIII] velocity offset 
larger than 50 \kms. Using a much smaller sample of 20 Seyfert 1 galaxies 
based on the observations of \citet{nw95}, \citet{cr10} reported that 50\% of 
Seyfert 1 galaxies have the [\OIII] velocity offset larger than 50 \kms. 

In the case of type 2 AGNs, \citet{cr10} presented that 17 out of 45 (38\%) 
Seyfert 2 galaxies have the velocity offset larger than 50 \kms\ while \citet{wa11} 
showed that $\sim$25\% of $\sim$3,000 type 2 AGNs from SDSS have the [\OIII] velocity offset 
above 50 \kms. Note that the velocity offset in the study of \citet{wa11}  was measured with respect 
to the low-ionization lines.
If we measure the [\OIII] velocity offset with respect to the \Ha\ lines using our AGN sample, 
$\sim$9\% of AGNs show the velocity offset $\ge$ 50 \kms\ (see Figure 10), which is lower than that of \cite{wa11}. 
One of the possible explanations for the difference is that their sample is biased toward the AGNs with high [\OIII] fluxes due to their selection 
with S/N $\ge$ 30 for [\OIII], while our sample includes AGNs with much weaker [\OIII].

Compared to type 1 AGNs, the outflow fraction of type 2 AGNs is generally lower for a fixed velocity offset limit (i.e., 50 \kms).
This can be explained as the orientation effect \citep{am85}  since
the projected outflow velocity to the line-of-sight is on average lower  in type 2 AGNs than 
in type 1 AGNs. The direct comparison of the outflow fraction between type 1 and type 2 AGNs
is beyond the scope of this study since the consistent measurements of the [\OIII] 
velocity offsets with respect to the systemic velocity (e.g., based on stellar absorption lines)
are also required for type 1 AGNs.

\begin{figure}
\centering
\includegraphics[width=0.49\textwidth]{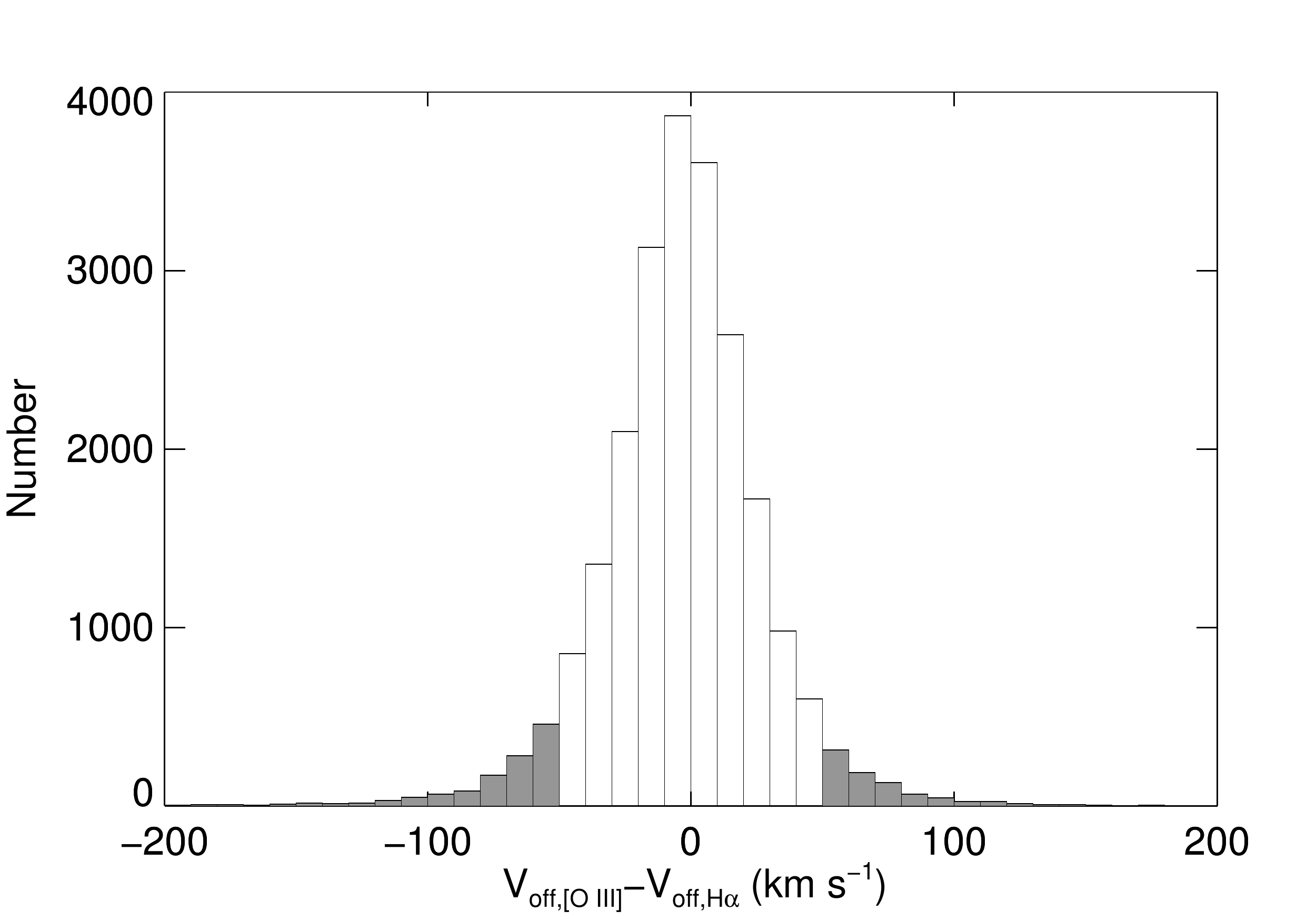}
\caption{Distributiuon of the velocity difference between [\OIII] and H$\alpha$. The shaded region indicates the AGNs with the velocity offset $\ge$ 50 \kms.}
\end{figure}

\section{Summary \& Conclusion}

Using a large sample of $\sim$23,000 type 2 AGNs, we perform a census of the ionized gas outflows 
by measuring the velocity offset of narrow emission lines with respect to 
the systemic velocity measured from the stellar absorption lines.  
We summarize the main results as follows.

\medskip

$\bullet$ We find that $\sim$47\% of type 2 AGNs display an [\OIII]
line-of-sight velocity offset larger than 20 \kms\
with respect to the systemic velocity.

$\bullet$ 
AGNs with larger [\OIII] velocity offsets, in particular with no or weak \Ha\ 
velocity offsets, tend to have higher Eddington ratios, 
implying 
that the [\OIII] velocity offset is related to on-going black hole activity.
The presence of high Eddington ratio AGNs with relatively small [\OIII] velocity 
offsets implies a projection effect, which can significantly decreases 
the line-of-sight velocity if the outflow direction is close 
to the plane of the sky.

$\bullet$ Face-on galaxies preferentially host the AGNs with blueshifted [\OIII] (rather than the AGNs with redshifted [\OIII]), 
presumably due to the dust obscuration in the host galaxy disk. 
The difference of the galaxy inclinations
between the AGNs with blueshifted and the AGNs with redshifted [\OIII] lines, supports
the combined model of the biconical outflow and dust obscuration.

$\bullet$ The outflow fraction in type 2 AGNs is a lower limit,
considering the detection limit of the velocity offset and the projection effect. 
For a fixed limit of the velocity offset (i.e., $>$ 50 \kms), type 2 AGNs have a lower outflow fraction than type 1 AGNs due to the projection.

$\bullet$ We find that for $\sim$3\% of AGNs, \Ha\ and [\OIII] show comparable velocity offsets, suggesting non-decelerating outflows and/or more complex gas kinematics in the NLR geometry, e.g., off-centered AGNs. 

\medskip

Based on these results, we conclude that the ionized gas outflows are common in AGNs.
The sample of the AGN-driven outflows with a large [\OIII] velocity
is one of the best suites for studying 
AGN feedback and the role of AGNs in the coevolution of SMBHs and their host galaxies.
Follow-up observations with high spatial resolution will provide detailed information on 
the kinematics and geometry of the NLR as well as the properties of ISM and the stellar 
populations in the host galaxies.

\acknowledgments
We thank the anonymous referee for valuable comments, which improved the presentation 
and clarity of the paper.
The work of HJB was supported by NRF (National Research Foundation of Korea) Grant funded by the Korean Government (NRF-2010-Fostering Core Leaders of the Future Basic Science Program).
JHW acknowledges the support by the National Research Foundation of Korea (NRF) grant funded 
by the Korea government (No. 2012-006087). 
Funding for the SDSS and SDSS-II has been provided by the Alfred P. Sloan Foundation, the Participating Institutions, the National Science Foundation, the U.S. Department of Energy, the National Aeronautics and Space Administration, the Japanese Monbukagakusho, the Max Planck Society, and the Higher Education Funding Council for England. The SDSS Web Site is http://www.sdss.org/.

\bibliographystyle{apj}

%http://merkel.zoneo.net/Latex/natbib.php
%\citet{jon90}	    		-->    	Jones et al. (1990)
%\citet[chap. 2]{jon90}	   	-->    	Jones et al. (1990, chap. 2)
%\citep{jon90}	   		-->    	(Jones et al., 1990)
%\citep[chap. 2]{jon90]	-->    	(Jones et al., 1990, chap. 2)
%\citep[see][]{jon90}	   	-->    	(see Jones et al., 1990)
%\citep[see][chap. 2]{jon90}	-->    	(see Jones et al., 1990, chap. 2)
%\citet*{jon90}	    		-->    	Jones, Baker, and Williams (1990)
%\citep*{jon90}	    		-->    	(Jones, Baker, and Williams, 1990)
%\citet{jon90,jam91}	    	-->    	Jones et al. (1990); James et al. (1991)
%\citep{jon90,jam91}	    	-->    	(Jones et al., 1990; James et al. 1991)
%\citep{jon90,jon91}	    	-->    	(Jones et al., 1990, 1991)
%\citep{jon90a,jon90b} 	-->    	(Jones et al., 1990a,b)

\clearpage

\end{document}